\pdfoutput=1

\documentclass[aps,pra,10pt,twocolumn,showpacs,superscriptaddress,nofootinbib]{revtex4}
\usepackage{amsmath}
\usepackage{latexsym}
\usepackage{amssymb}
\usepackage{graphics,epstopdf}

\begin{document}

\title{Nonlocal continuous variable correlations and violation of Bell's inequality for light beams with topological singularities}

\author{Priyanka Chowdhury}

\affiliation{S. N. Bose National Centre for Basic Sciences, Block JD, Sector III, Salt Lake, Kolkata 700098, India}

\author{A. S. Majumdar}

\affiliation{S. N. Bose National Centre for Basic Sciences, Block JD, Sector III, Salt Lake, Kolkata 700098, India}

\author{G. S. Agarwal}

\affiliation{Department of Physics, Oklahoma State University, Stillwater, Oklahoma 74078, USA}

\begin{abstract}
We consider optical beams with topological singularities which possess Schmidt
decomposition and show that such classical beams share many features of two mode
entanglement in quantum optics. We demonstrate the coherence properties of such
beams through the violations of Bell inequality for continuous variables using 
the Wigner function. This violation is a consequence of correlations between 
the $(x, p_x)$ and $(y, p_y)$ spaces which mathematically play the same role as 
nonlocality in quantum mechanics. The Bell violation for the LG beams is shown 
to increase with higher orbital angular momenta $l$ of the vortex beam.  This 
increase is reminiscent of enhancement of nonlocality for many particle 
Greenberger-Horne-Zeilinger states or for higher spins. The states with 
large $l$ can be easily produced using 
spatial light modulators.


\end{abstract}

\date{\today}

\maketitle

\section{Introduction}

The realization that light can be twisted like a corkscrew around its axis of 
propagation, thus being endowed with interesting features such as topological 
charge and singular structure \cite{nye}, has lead to remarkable and
diverse applications in recent years, such as in optical tweezers \cite{tweezer}, 
and in the detection of exo-planets \cite{exoplanet}. This has also inspired 
deeper studies on the coherence properties  of vortex beams 
\cite{simon,banerji}.
Further, the possibility of encoding large amounts of information in
vortex beams due to the absence of an upper limit on their topological
charge and a corresponding number of allowed states, has raised the prospects 
of their applicability in quantum information processing tasks such as 
computation and cryptography \cite{review}. Vortex beams
with large values of orbital angular momenta have been experimentally realized 
both in the optical
domain \cite{fickler}, as well as using electrons \cite{science}.

Understanding the coherence properties of vortex beams is central to
manipulating them for various applications. It has been realized that traditional coherence measures may be inadequate when more than one
coupled degree of freedom of light is involved in a practical situation \cite{wolf}. A similar issue of coupled degrees of freedom has
though been much studied in the quantum domain in the form of
quantum entanglement \cite{horod}. Taking inspiration from the mathematical isomorphism of correlations between discrete degrees of freedom in
classical optics with quantum entanglement in two-qubit systems, a
common framework to study correlations in discretized degrees of freedom in both classical and quantum optics has been
proposed \cite{Aiello, simon2}. It may be noted that the Schmidt decomposition which has been known much
before the advent of quantum mechanics, plays a key role in defining quantum entanglement when one uses the
wave function. In classical theory the Schmidt decomposition is in terms of the electromagnetic
fields, and the coherence function is directly related to the Schmidt spectrum \cite{jha}. It is recognized
that LG beams that have topological singularities have a Schmidt decomposition \cite{banerji}, and therefore, one would expect that many of the ideas
developed within the context of quantum mechanics would be applicable to LG beams as well. In particular, Schmidt index
was used in \cite{banerji} to study the mixedness of the 
one dimensional projections of
the LG beams.

In the same spirit, Bell's measure  {\it viz.} the amount of violation of a Bell inequality \cite{bell,chsh},
has recently been suggested as a measure to quantify the magnitude of
correlation between degrees of freedom of a classical beam, through joint measurements \cite{saleh}. It is worth emphasizing that in
the derivation of Bell's inequality probability theory is invoked, with quantum mechanics playing no role, and thus, Bell's measure is not
exclusively related to quantum phenomena. The physical significance of the violation of Bell's inequality in quantum mechanics, {\it viz.} quantum nonlocality \cite{bell,bohm},
is reinterpreted in classical theory where a violation corresponding to a
particular light beam possessing classical correlations signifies the impossibility of constructing such a beam using other beams with
uncoupled degrees of freedom. Such quantum inspired inseparability with several interesting consequences, has been variously dubbed in the literature as `nonquantum entanglement' \cite{simon2} or `classical entanglement'
\cite{saleh}. Moreover, the Schmidt decomposition for LG beams contains many more terms for larger values of
the orbital angular momentum \cite{banerji}, and hence, we  expect a 
rise in the Bell measure for higher orbital angular momentum.

We note that the quantum entanglement is quite natural to two particle quantum mechanics. It is a 
consequence of the superposition principle which allows us to write the two particle wave function
$\psi(x,y)$ as a superposition of the product of the single particle wave functions $\Phi_i(x)$,
$\chi_i(y)$,
\begin{eqnarray}
\psi(x,y) = \sum_i c_i \Phi_i(x) \chi_i(y)
\label{super1}
\end{eqnarray}
This is a nonseparable state as long as there are at least two nonzero $c_i$'s. In fact, the state
(\ref{super1}) is in the form of Schmidt decomposition which has been extensively used in studying
quantum entanglement \cite{eberly,sper1}.  A way to study entanglement 
is to study the nonpositivity
of the quasi probabilities \cite{sper2}. For the classical fields with topological
singularities considered in the present paper there is a close parallel to the development in quantum
mechanics. It is well known that in paraxial optics, the beam propagation in free space is described by 
\cite{saleh2} 
\begin{eqnarray}
E(\vec{r},t) = \varepsilon(x,y,z) e^{i\omega z/c - i\omega t},
\label{fr1}
\end{eqnarray}
\begin{eqnarray}
i \frac{\partial \varepsilon}{\partial z} = -\frac{\lambdabar}{2} \left( \frac{\partial^2}{\partial x^2}
+ \frac{\partial^2}{\partial y^2}\right) \varepsilon
\label{freespace}
\end{eqnarray}
with $\lambdabar = \lambda/2\pi$, $2\pi/\lambda = \omega/c$. Eq.(\ref{freespace}) has exactly the same
form as the Schrodinger equation for a free particle in two dimensions with $t \to z$, $\psi \to \varepsilon$,
$\hbar \to \lambdabar$. Thus, an optical beam in two dimensions can be expressed as a superposition of
fundamental solutions of Eq.(\ref{freespace}). For example, the well known LG beam in two dimensions,
which is a physically realizable field distribution
containing optical vortices with topological singularities is given by \cite{book}
\begin{eqnarray}
\Phi_{nm}(\rho,\theta) = e^{i(n-m)\theta}e^{-\rho^2/w^2}(-1)^{\mathrm{min}(n,m)}
\left(\frac{\rho \sqrt{2}}{w}\right)^{|n-m|} \nonumber \\
\times \sqrt{\frac{2}{\pi n! m ! w^2}}
L^{|n-m|}_{\mathrm{min}(n,m)} \left(\frac{2\rho^2}{w^2}\right) (\mathrm{min}(n,m)) !
\label{waveLG}
\end{eqnarray}
with
 $\int |\Phi_{nm}(\rho,\theta)|^2 dx dy =1$,
where $w$ is the beam waist, and $L_p^l(x)$ is the generalized Laguerre polynomial. These LG beams can be written as superpositions
of Hermite-Gaussian (HG) beams \cite{danakas}
\begin{eqnarray}
\Phi_{nm}(\rho,\theta) = \sum_{k=0}^{n+m} u_{n+m-k,k}(x,y)\frac{f_k^{(n,m)}}{k!}(\sqrt{-1})^k \nonumber \\
\times \sqrt{\frac{k! (n+m-k)!}{n! m! 2^{n+m}}}
\label{legherm}
\end{eqnarray}
\begin{eqnarray}
f_k^{(n,m)} = \frac{d^k}{dt^k} ((1-t)^n(1+t)^m)|_{t=0},
\label{def11}
\end{eqnarray}
and the HG beam is defined by
\begin{eqnarray}
u_{nm}(x,y) = \sqrt{\frac{2}{\pi}} \left(\frac{1}{2^{n+m} w^2 n!m!}\right)^{1/2} \nonumber\\
\times H_n \left(\frac{\sqrt{2}x}{w}\right) H_m \left(\frac{\sqrt{2}y}{w}\right) e^{-(x^2+y^2)/w^2},
\nonumber \\
 \int |u_{nm}(x,y)|^2 dx dy =1
\label{hermite}
\end{eqnarray}
The superposition (\ref{legherm}) is like a Schmidt decomposition. 
In the special case
\begin{eqnarray}
\Phi_{10} = \sqrt{\frac{2}{\pi w^2}} (x + iy)e^{-(x^2+y^2)/w^2} \nonumber\\
\Phi_{01} = \sqrt{\frac{2}{\pi w^2}} (x - iy)e^{-(x^2+y^2)/w^2}
\label{zerolg} 
\end{eqnarray}

Motivated by the structural similarities between  Eq.(\ref{super1}) for a quantum system and  Eq.(\ref{legherm}) for optical beams, we examine the possibilities of violations of Bell like inequalities for classical optical
LG beams. We indeed show that Bell like inequalities are violated for such beams. We also analyze the reasons
for such violations. We show how the correlations between $(x, p_x)$  and
$(y, p_y)$ spaces are responsible for violations of Bell like
inequalities. The paper is organized as follows. In the next Section we briefly discuss the framework of obtaining Bell
inequalities for continuous variable systems using the Wigner function. In Section III we present 
the violation the Bell-CHSH inequality by LG beams. Here we also show how this violation increases 
with the increase of orbital angular momentum of the beam. In Section IV we present a further explanation
of this violation in terms of the two-mode correlation function that is shown to exhibit a similar
increase of magnitude with orbital angular momentum. The last Section is reserved for a summary and concluding remarks.

 \section{Bell inequalities for continuous variable systems}
 
 In local hidden variable theories the correlations between the outcomes of measurements on two
spatially separated systems with detector settings labeled by $\mathbf{a}$ and $\mathbf{b}$, respectively,
may be written as a statistical average over hidden variables $\tau$, of the functions $p(\mathbf{a}, \tau) = \pm 1$, and $p(\mathbf{b}, \tau) = \pm 1$, {\it viz.}
\begin{eqnarray}
S(\mathbf{a}, \mathbf{b}) = \int d \tau \rho(\tau) p(\mathbf{a}, \tau) p(\mathbf{b}, \tau)
\label{bellin}
\end{eqnarray}
where $\rho(\tau)$ is a local and positive distribution of the hidden variables $\tau$.
Using the choice of two different settings ($\mathbf{a}, \mathbf{a'}, \mathbf{b}, \mathbf{b'}$) on either side, the Bell-CHSH inequality \cite{bell,chsh}, {\it viz.},
\begin{eqnarray}
B \equiv |S(\mathbf{a}, \mathbf{b}), + S(\mathbf{a}, \mathbf{b'}) + S(\mathbf{a'}, \mathbf{b}) - S(\mathbf{a'}, \mathbf{b'})| < 2
\label{Bel}
\end{eqnarray}
may be derived, which has been shown to be violated experimentally for
quantum systems with correlations in discrete variables \cite{aspect}.

 The entanglement in quantum systems with continuous variables (non `qubit' 
systems) is usually
 characterized in terms of the quasiprobabilities. For example, Banaszek and 
Wodkiewicz \cite{banaszek}
 have argued that the Wigner function expressed as an expectation value of a 
product of
displaced parity operators,
can be used to derive an analog of Bell inequalities in continuous
 variable systems. There exists an analogy between the measurement of 
spin-$1/2$ projectors
and the parity operator, since the outcome of a measurement of the latter is 
also dichotomic. The
solid angle defining the direction of the spin in the former case, is 
replaced by the coherent displacement
describing the shift in phase space in the latter.
For a radiation field with two modes $a$ and $b$, we 
replace $S(\mathbf{a}, \mathbf{b})$   
 in Eq.(\ref{bellin}) by the function
 $W(\alpha,\beta) = \frac{4}{\pi^2} \langle D_1(\alpha)(-1)^{a^{\dagger}a}D_1^{\dagger}(\alpha) \otimes
 D_2(\beta)(-1)^{b^{\dagger}b}D_2^{\dagger}(\beta) \rangle$
 where $D(\alpha) = exp{\{a^{\dagger}\alpha - a\alpha^{*}\}}, [a,a^{\dagger}]=1=[b,b^{\dagger}]$.
 Thus, for continuous variable systems, one can test the violations of the 
inequality
 \begin{eqnarray}
 B = && \frac{\pi^2}{4}|W(\alpha,\beta)+ W(\alpha,\beta')+ W(\alpha',\beta)\nonumber \\
 && - W(\alpha',\beta')| < 2.
 \label{belin12}
 \end{eqnarray}
The approach of using the Wigner function for demonstrating
the violation of Bell inequalities for continuous variables in quantum optics has gained popularity in recent years \cite{zhang,paris,zhang2}.
 We will use the inequality (\ref{belin12}) for classical light beams to find the features of quantum
 inspired optical entanglement. 

\section{Violation of Bell's inequality through the Wigner function}

Defining the Wigner function  as
 the Fourier transform
of the electric field amplitude $E$, {\it viz.}
\begin{eqnarray}
W(\mathbf{X},\mathbf{P}) = \frac{1}{\pi^2} \int d^2 \mathbf{\xi} e^{2i \mathbf{P} \mathbf{\xi}}
\langle E^{*}(\mathbf{X} - \mathbf{\xi}) E(\mathbf{X} - \mathbf{\xi}) \rangle
\label{wigel}
\end{eqnarray}
has facilitated the experimental measurement of the Wigner function in terms of the two-point
field correlations \cite{zhang,iaconis}. The Wigner function 
 has been calculated  for LG beams \cite{simon}:
 \begin{eqnarray}
 W_{nm}(x,p_x;y,p_y) =  (-1)^{n+m}(\pi)^{-2}L_{n}[4(Q_0+Q_2)]\nonumber \\
L_{m}[4(Q_0-Q_2)]~exp(-4Q_0)
\label{WF_LG_n1_n2}
\end{eqnarray}
where the  expressions for $ Q_0 $ and $ Q_2 $ are given as follows
\begin{eqnarray}
Q_0 & = &  \frac{1}{2} \left[ \frac{x^2 + y^2}{w^2} + \frac{w^2}{4 \lambdabar^2} (p_x^2+p_y^2) \right] \nonumber \\
Q_2 & = &  \frac{x p_y - y p_x}{2 \lambdabar}.
\end{eqnarray}

Before proceeding further, we make the following variable transformations,
 \begin{eqnarray}
x (y) \rightarrow \frac{w}{\sqrt{2}} ~~X (Y),
\hspace{0.5cm}
p_x (p_y) \rightarrow \frac{\sqrt{2} \lambdabar}{w} ~~P_X (P_Y),
\label{DL_T}
\end{eqnarray}
 where $\{X,~P_X\}$ and $\{Y,~P_Y\}$ are conjugate pairs of dimensionless quadratures. With the above transformation, $[\hat{x},\hat{p}_x]=i\lambdabar;~[\hat{y},\hat{p}_y]=i\lambdabar$ becomes $[\hat{X},\hat{P}_X]=i;~~[\hat{Y},\hat{P}_Y]=i$, and the operators $\hat{P}_X$ and $\hat{P}_Y$ are given by
\begin{eqnarray}
\hat{P}_X = - i \frac{\partial}{\partial X},
\hspace{0.5cm}
\hat{P}_Y = - i \frac{\partial}{\partial Y}
\label{M_OP}
\end{eqnarray}
The Wigner function is rewritten  in terms of the scaled 
variables as
\begin{eqnarray}
W_{nm}(X,P_X;Y,P_Y) &=& (-1)^{n+m}(\pi)^{-2}L_{n}[4(Q_0+Q_2)]\nonumber \\
L_{m}[4(Q_0-Q_2)]~exp(-4Q_0), \nonumber \\
 Q_0 &=& \frac{1}{4}\left[ X^2 + Y^2 + P_X^2+P_Y^2\right], \nonumber\\
Q_2 &=& \frac{XP_Y-YP_X}{2}
\label{WF_LG_n1_n2}
\end{eqnarray}
with the normalization $\int W_{nm}(X,P_X;Y,P_Y) dX dY dP_X dP_Y =1$.
In terms of the Wigner function for the LG beam, we would search violations of the analog of Eq.(\ref{belin12}) given by 
\begin{eqnarray}
B&=& \Pi(X=0,P_{X}=0;Y=0,P_{Y}=0)
 +\Pi(X,0;0,0)  \nonumber\\ &+& \Pi(0,0;0,P_{Y}) - \Pi(X,0;0,P_{Y}) < 2
\label{Bel2}
\end{eqnarray}
where the Wigner transform $\Pi_{nm}$ \cite{zhang2} associated with $W_{nm}(X,P_X;Y,P_Y)$  is given by
\begin{equation}
\Pi_{nm}(X,P_X;Y,P_Y))=(\pi)^2 \; W_{nm}(X,P_X;Y,P_Y).
\label{WT_LG_n1_n2}
\end{equation}

We again emphasize that since the expression for correlations in joint measurement of separated observables given by Eq.(\ref{bellin})
is not exclusive to the quantum domain, the above formulation of Bell inequalities through the
Wigner function may also be
applied in classical theory. Note that a Bell inequality
 involving correlations between the discrete variables of polarization and parity has been shown
 to be violated in classical optics \cite{saleh}.
In our present analysis we apply the framework of the Wigner function formulation of the Bell-CHSH
inequality for the first time in classical optics to study the continuous variable correlations in light beams
with topological singularities. In particular, we apply the above framework to the case of LG beams.

\subsection{Bell violation for $n=1$, $m=0$}

Let us first consider the state $\Phi_{10}(X,Y)$ (given by Eq.(\ref{zerolg})) 
which in terms of the variables $(X,Y)$ is given by
\begin{eqnarray}
\Phi_{10}(X,Y)= \frac{1}{\sqrt{\pi}} (X + i Y) \exp[-\frac{{X}^2+{Y}^2}{2}],
\label{waveLG_1_2_DL}
\end{eqnarray}
The corresponding normalized Wigner function  is given by
\begin{eqnarray}
W_{10}(X,P_X;Y,P_Y)
&=& e^{-P_X^2-P_Y^2-X^2-Y^2}  \\
&\times&   \frac{ \left( (P_X - Y)^2 +(P_Y+X)^2 -1\right)}{\pi^2} \nonumber
\end{eqnarray}
 In order to obtain the Bell sum, we consider the Wigner transform \cite{zhang2}
$\Pi_{10}(X,P_X;Y,P_Y) = \pi^2 W_{10}(X,P_X;Y,P_Y)$.
 The two measurement settings on one side are chosen to be  $\{X1=0,P_{X1}=0\}$ or  $\{X2=X,P_{X2}=0\}$, and the corresponding settings on the other side are  $\{Y1=0,P_{Y1}=0\}$ or  $\{Y2=0,P_{Y2}=P_Y\}$ \cite{zhang2}. Hence, the Bell sum associated with  $ \Pi_{10}(X,P_X;Y,P_Y) $ for the bimodal state $\Phi_{10}(X,Y)$ is given by
\begin{eqnarray}
B&=& \Pi_{10}(X=0,P_{X}=0;Y=0,P_{Y}=0)\nonumber \\
&& +\Pi_{10}(X,0;0,0)+\Pi_{10}(0,0;0,P_{Y}) - \Pi_{10}(X,0;0,P_{Y}) \nonumber \\
&=& e^{-P_Y^2} \left(P_Y^2-1\right) + e^{-X^2}\left(X^2-1\right) \nonumber \\
&-& e^{-P_Y^2-X^2} \left((P_Y+X)^2-1\right) -1
\label{BI_1_3}
\end{eqnarray}
 Upon maximization of the Bell sum $B$ with respect to parameters $X$ and $P_{Y}$, we obtain the maximum
 Bell violation, $|B_{\max}| \sim 2.17$ which occurs for the choices of parameters $X \sim 0.45,~P_{Y} \sim 0.45$. Note here for comparison that the maximum Bell violation in quantum mechanics through the Wigner function for the two-mode squeezed vacuum state using similar settings is given
 by $|B_{\max}|_{QM} \sim 2.19$ \cite{banaszek}.

 \subsection{Bell violation for higher values of $n$, $m$}

\begin{figure}[!ht]
\resizebox{9cm}{4cm}{\includegraphics{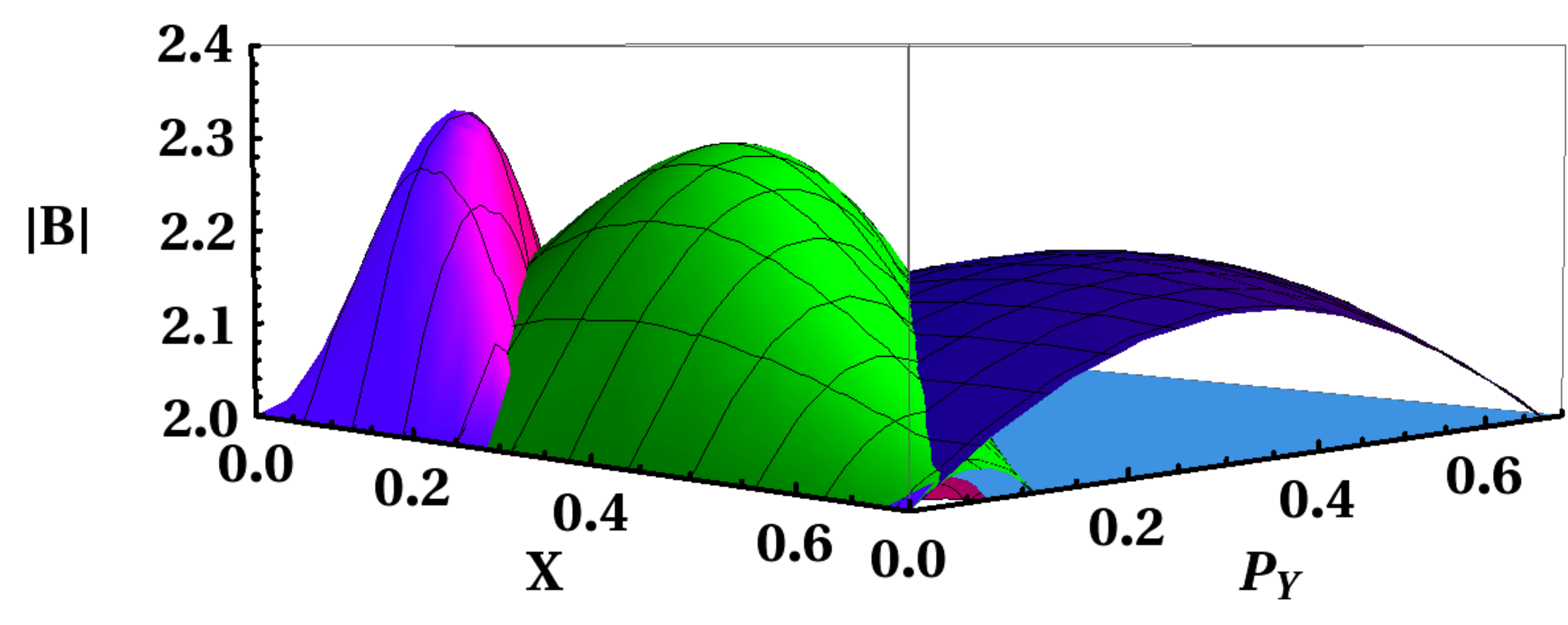}}
\caption{\footnotesize (Coloronline) The plot shows the variation of the Bell sum $|B|$ with respect to dimensionless variables $X$ and $P_Y$ for different values of $n$, where $m=0$. The indigo (right) curve is for $n=1$, the green (centre) curve is for  $n=5$, and the magenta (left) curve is for $n=30$.
}
\label{FIGBI}
\end{figure}

 We next repeat the above analysis for higher values of $n$ and $m$ for the LG field amplitude. We use  Eqs.(\ref{WF_LG_n1_n2}) and (\ref{WT_LG_n1_n2}) to calculate the Bell sum.
 In the Figure~\ref{FIGBI}, we plot $|B|$ against $X$ and $P_Y$ for three different values of $n$ keeping
 $m=0$. We find that the violation of the Bell's inequality increases with higher orbital angular momentum.
 The increase of Bell violations with $n$ is analogous to the enhancement of 
nonlocality in quantum
 mechanics for many particle Greenberger-Horne-Zeilinger states or for
higher spins \cite{mermin}, an effect which may also be manifested
 in physical situations \cite{nayak}. Here we have been able to demonstrate such an effect within the
 realm of classical theory.

 We would like to note here that for the purpose of experimental realization of the violation of Bell inequalities in classical optical
 systems with topological singularities, it may be worthwhile to employ techniques for enhancing the
 Bell violation. This is indeed possible using several approaches, and we would here like to point out
 two such schemes. First, it has been observed \cite{paris} that the Bell violation may be further optimized by a more general choice of settings than those used by
 us in obtaining the Bell sum, i.e.,
 \begin{eqnarray}
B &=& \Pi_{j,m}^{LG}(X1,P_{X1};Y1,P_{Y1}))+\Pi_{j,m}^{LG}(X2,P_{X2};Y1,P_{Y1}))\nonumber \\
 &&+\Pi_{j,m}^{LG}(X1,P_{X1};Y2,P_{Y2}))\nonumber \\
 && -\Pi_{j,m}^{LG}(X2,P_{X2};Y2,P_{Y2}))
\label{BInequality}
\end{eqnarray}
 Considering the $n=1, m=0$ case, and maximizing the Bell violation with respect to the  parameters $X1,P_{X1},X2,P_{X2},Y1,P_{Y1},Y2,P_{Y2}$,
 one obtains the maximum Bell violation, $|B_{\max}| = 2.24$ which exceeds the maximum violation obtained
 through our earlier choice of settings given by Eq.(\ref{BI_1_3}), and occurs for the choices of parameters $X1 \sim -0.07,~P_{X1} \sim 0.05,~X2 \sim 0.4,~P_{X2} \sim -0.26,~Y1 \sim-0.05,~P_{Y1} \sim -0.07,~Y2 \sim 0.26,~P_{Y2} \sim 0.4$. Similarly, a corresponding increase of the Bell sum occurs for higher values of
$n$ too. Secondly, another method of obtaining higher violation of Bell inequalities may be through
elliptical transformations of LG beams. Such transformations are easily achievable
in practice \cite{gaussian}, {\it viz.} a Gaussian elliptical beam of the sort
\begin{eqnarray}
\Phi = \frac{1}{\sqrt{\pi}} \exp[-(X^2+Y^2)/2 \cosh{2t} \pm XY \sinh(2t)
\label{ellipse}
\end{eqnarray}
is observed to increase the Bell violation for the $n=1, m=0$ case to $2.32$.

\section{Nonlocal correlations and Bell violations in vortex beams}

The violation of the Bell's inequality obtained above follow from nonvanishing correlations
between the two modes of the type $<X,P_Y> \neq 0$, with $<X>=0=<P_Y>$ individually.
In quantum mechanics  the correlations between two non-commuting
observables of a sub-system with those of the other sub-system have rich consequences. In wave optics
the wavelength $\lambdabar$ plays a role analogous
to the Planck's constant $\hbar$ in quantum mechanics. Thus, nonlocal
correlations of the type $<X,P_Y> \neq 0$ originate due to
the finite and non-vanishing wavelength $\lambdabar$, resulting in
the lack of precision in simultaneous measurement of two observables
corresponding to two different modes of light. Note that the above correlations are between
separate modes or variables in separate directions, {\it viz.}, position 
in the $x$-direction,
and momentum in the $y$-direction. Here
 $\lambdabar \to 0$ leads to the limit of geometrical optics, again
 analogously to the quantum case where $\hbar \to 0$  gives the classical
 limit.

 Let us now consider the situation where the quadrature phase components of two correlated and spatially
separated light fields are measured.  The quadrature amplitudes associated with the fields $E_{\alpha}=C[\hat{\alpha} e^{-i\omega_{\alpha} t} + \hat{\alpha}^{\dagger} e^{i\omega_{\alpha} t}]$ (where, $\alpha\in\{a,b\}$, are the bosonic operators for two different modes, $\omega_{\alpha}$ is the frequency, and
$C$ is a constant incorporating spatial factors taken to be equal for each mode) are given by
\begin{eqnarray}
\hat{X}_{\theta}=\frac{\hat{a}e^{- i \theta} + \hat{a}^{\dagger} e^{i \theta}}{\sqrt{2}},
\hspace{0.5cm}
\hat{Y}_{\phi}=\frac{\hat{b}e^{- i \phi} + \hat{b}^{\dagger} e^{i \phi}}{\sqrt{2}},
\label{Quard}
\end{eqnarray}
where,
\begin{eqnarray}
\hat{a} &=& \frac{X + i P_x}{\sqrt{2}},\hspace{0.5cm} \hat{a}^\dagger = \frac{X -i P_x}{\sqrt{2}},\nonumber\\
\hat{b}&=&  \frac{Y+i P_y}{\sqrt{2}}, \hspace{0.5cm} \hat{b}^\dagger = \frac{Y- i P_y}{\sqrt{2}},
\label{boson_op}
\end{eqnarray}
and the commutation relations of the bosonic operators are given by $[\hat{a},\hat{a}^{\dagger}]=1=[\hat{b},\hat{b}^{\dagger}]$.
Now, using Eq.(\ref{boson_op})  the expression for the quadratures can be rewritten as
\begin{eqnarray}
\hat{X}_{\theta} = \cos[\theta] ~\hat{X} + \sin[\theta] ~\hat{P}_x, \hspace{0.5cm}
\hat{Y}_{\phi} = \cos[\phi]~ \hat{Y} +\sin[\phi]~ \hat{P}_y.
\label{Dless}
\end{eqnarray}
The correlations between the quadrature amplitudes $\hat{X}_{\theta}$ and $\hat{Y}_{\phi}$ are captured by the correlation coefficient, $ C_{\theta,\phi} $  defined as \cite{reid,ou,tara}
\begin{eqnarray}
C_{\theta,\phi}=\frac{\langle \hat{X}_{\theta} \hat{Y}_{\phi} \rangle}{\sqrt{\langle \hat{X}^2_{\theta} \rangle \langle \hat{Y}^2_{\phi}  \rangle}},
\label{Cr_f}
\end{eqnarray}
where $\langle \hat{X}_{\theta} \rangle=0=\langle \hat{Y}_{\phi} \rangle$. The correlation is perfect for some values of $\theta$ and $\phi$, if $|C_{\theta,\phi}|=1$. Clearly $|C_{\theta,\phi}|=0$ for uncorrelated variables. For the case of LG beams with $n=1,m=0$, the
correlation function is given by
\begin{eqnarray}
C_{\theta,\phi} (\Phi_{10}(X,Y))= \frac{1}{2} \sin[\phi-\theta],
\label{Cor_n1_1}
\end{eqnarray}
Here, the maximum correlation strength $|C^{\max}_{\theta,\phi}| ~=\frac{1}{2}$ occurs for $\phi-\theta=\frac{k \pi}{2} $ (where $k$ is an odd integer). For arbitrary values of $n,m$ it can be shown that the
expression for the maximum correlation function is given by
\begin{eqnarray}
C^{\max}_{\theta,\phi} = \frac{\langle XP_Y \rangle}{\sqrt{\langle X^2 \rangle \langle P^2_Y\rangle}} =
- \frac{\langle P_X Y \rangle}{\sqrt{\langle P_X^2 \rangle \langle Y^2\rangle}}
\label{maxcor}
\end{eqnarray}

\vskip 0.2in

\begin{figure}[!ht]
\resizebox{9cm}{5cm}{\includegraphics{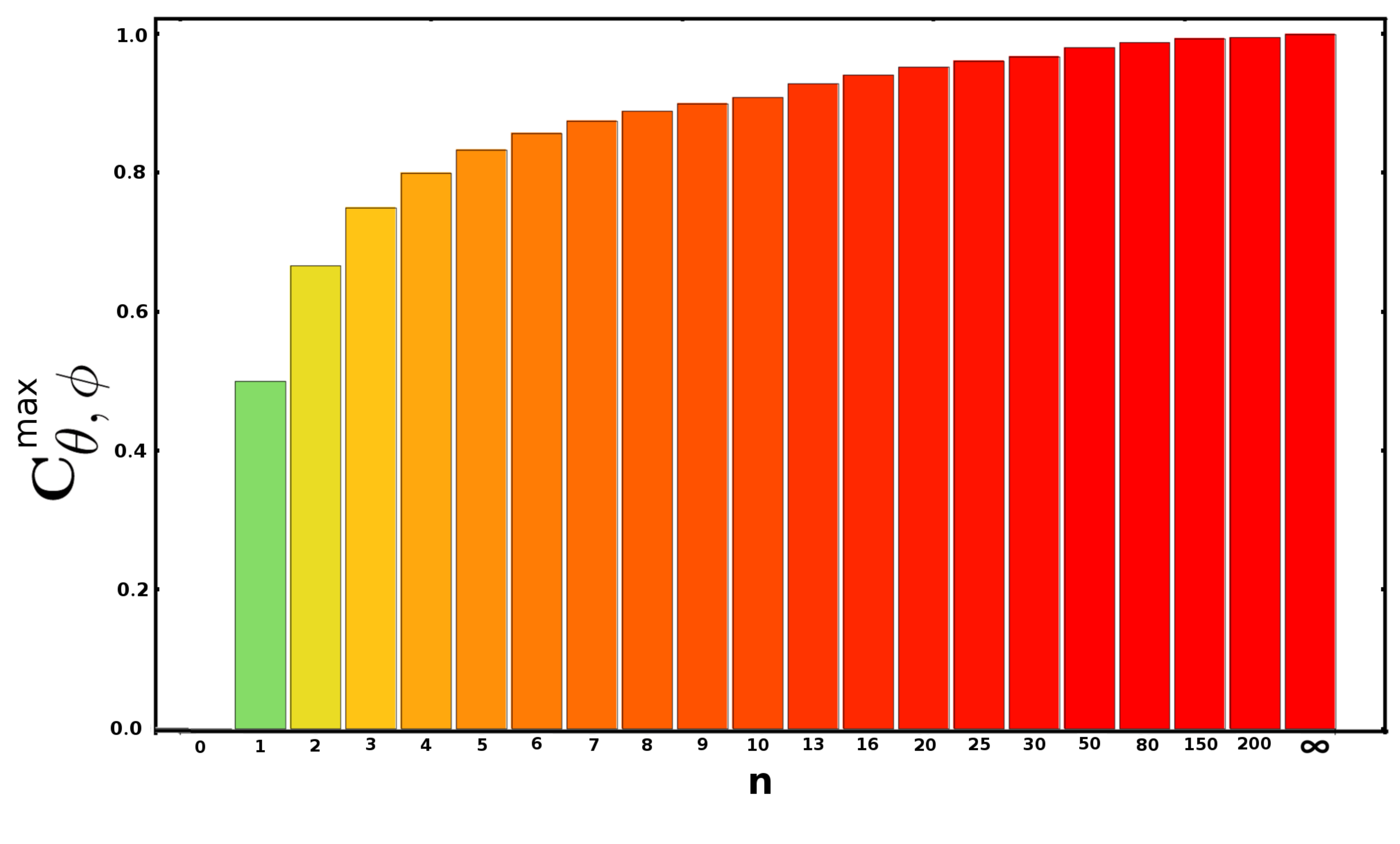}}
\caption{\footnotesize  (Coloronline) The plot  shows the values of the maximum correlation function $C^{\max}_{\theta,\phi}$  for
 various values of $n$, where $m=0$. Similar results are obtained by choosing
$n=0$ and varying $m$. Note that $C_{\theta,\phi}=0$ for $n=m=0$.
}
\label{Fig2}
\end{figure}

 In Fig.2 we provide a plot of the
the maximum correlation function for several values of $n,m$.  The strength of
the correlations increases with $n(m)$, asymptotically reaching the limit of perfect correlations
as $n$ becomes very large, as is expected to be the case due to the presence of more and more  terms
in the Schmidt decomposition of LG beams \cite{banerji}. This feature thus further corroborates  our earlier
results of increase in Bell violations for larger orbital angular momentum
of LG beams.

\vskip 0.2in

\section{Conclusions}

To summarize, in this work we have presented the first study of nonlocal correlations in classical optical
beams with topological singularities.  These nonlocal correlations between two different light modes
 are  manifested through the violation of a Bell inequality using the Wigner function for this system of
 classical vortex beams. We need to use the Wigner function as we are dealing
with two continuous variables. The magnitude of violation of the Bell inequality is shown to increase with
 the value of orbital angular momentum of the beam, an effect that is analogous to the enhancement of
 nonlocality for many particle Greenberger-Horne-Zeilinger states or for
higher spins \cite{mermin}. This feature is further corroborated by the corresponding
 increase of the quadrature correlation function. Our predicted values of the correlation function
 as function of the beam parameters should be not difficult to realize experimentally, since production
 of such vortex beams have been achieved not only in the optical domain \cite{fickler,singh}, but recently has
 also been implemented for electron beams \cite{science} having far-reaching applications. The feasibility of direct measurement of the two-point correlation function through shear Sagnac interferometry \cite{iaconis,zhang,singh} is a potentially promising avenue for experimental verification of
 our predicted Bell violation and its enhancement for vortex beams with higher angular momentum. We expect the results of this paper to hold also for other types
of beams with no azimuthal symmetry. An example would be Bessel 
beams \cite{bessel} of higher order $(J_l(\rho)e^{il\theta}; l\neq 0)$. As
emphasized in Section IV, we need nonlocal correlations, i.e., 
$\langle x, p_y \rangle \neq 0$, and beams with no azimuthal symmetry do 
have this property.

Clearly, the violation of the Bell inequality (\ref{Bel2}) for classical 
light fields and the existence of nonlocal correlations (\ref{Cor_n1_1}) 
bring out totally new statistical features of the optical beams. 
Traditionally, statistical optics is pursued in terms of the coherence
function defined as $\langle E^{*}(\vec{r},\omega)E(\vec{r}',\omega)\rangle$.
Here the brackets refer to the ensemble average. The new features are
contained in the quantities defined by (\ref{Cr_f}). The correlations like
$\langle x^2 \rangle$, $\langle y^2 \rangle$ give the standard beam
characteristics, whereas a correlation like $\langle x, p_y \rangle$ is
a correlation between two conjugate variables and can be studied by
examining fields in position and momentum spaces. The Wigner function
(\ref{WF_LG_n1_n2}) of the LG beams captures this aspect nicely via 
its dependence on the variable $Q_2$. Clearly, the present work provides
a new paradigm to the well developed optical coherence theory.

{\it Acknowledgements:} One of us (ASM) thanks Oklahoma State University for
the hospitality while this work was done.

\end{document}